\documentclass[final,2p,times,onecolumn]{article}

\usepackage[
  a4paper,
  left=2.0cm,
  right=2.0cm,
  top=2cm,
  bottom=2cm
]{geometry}

\usepackage{bibunits}
\usepackage{hyperref}
\usepackage{amssymb}
\usepackage{graphicx}
\usepackage{amsmath}
\usepackage{bm}
\usepackage{xcolor}
\usepackage[switch,mathlines]{lineno}
\usepackage{float}

\usepackage{xcolor}

\title{From Tunnel Overburden Reconstruction to Tidal Measurements via atmospheric muons}

\author{
Jiwon~Seo$^{1}$,
Jongseok~Chung$^{1}$,
Chang~Hyon~Ha$^{1}$,
TongZhou~Huang$^{1}$,
Jinyoung~Kim$^{1}$,\\
Sooa~Kim$^{1}$,
Hani~Kimku$^{1}$,
Byoung-cheol~Koh$^{1}$,
Minsu~Kwak$^{1}$,
Sihyun~Lee$^{1}$,
and Yujin~Lee$^{1}$\\[4pt]
\small $^{1}$Department of Physics, Chung-Ang University,
Seoul 06974, Republic of Korea\\
}

\date{}

\begin{document}

\begin{bibunit}[unsrt]

\maketitle

\begin{abstract}
  Muography with scintillator-based detectors enables the investigation of density distributions within large structures.
  The Hankuk Atmospheric-muon Wide Landscaping (HAWL) project is the first real-time portable muon tomography.
  It successfully charted the mountainous region of eastern Korea by measuring cosmic-ray muons with a detector operated in a moving vehicle.
  HAWL achieved a tunnel length accuracy of 6.0~\% and an overburden measurement range of 8-400 meter-water-equivalent (m.w.e.).
  The HAWL upgrade (HAWLU) features a more compact design and enables trajectory reconstruction.
  Using this upgraded system, we measured tidal water-level variations with meter-scale precision over 31.6 days in the Boryeong Undersea Tunnel.
  The measured muon rate decreased by $1.72 \pm 0.04$~\% per meter of increasing tidal level, while the reconstruction of undersea tunnel scan independently validated the detector performance.
  These results demonstrate the capabilities of dynamic muography in two approaches, one with a portable detector in motion and the other with a detector measuring a target that changes over time.
\end{abstract}

\begin{figure*}[!b]
  \begin{center}
      \includegraphics[width=0.90\textwidth]{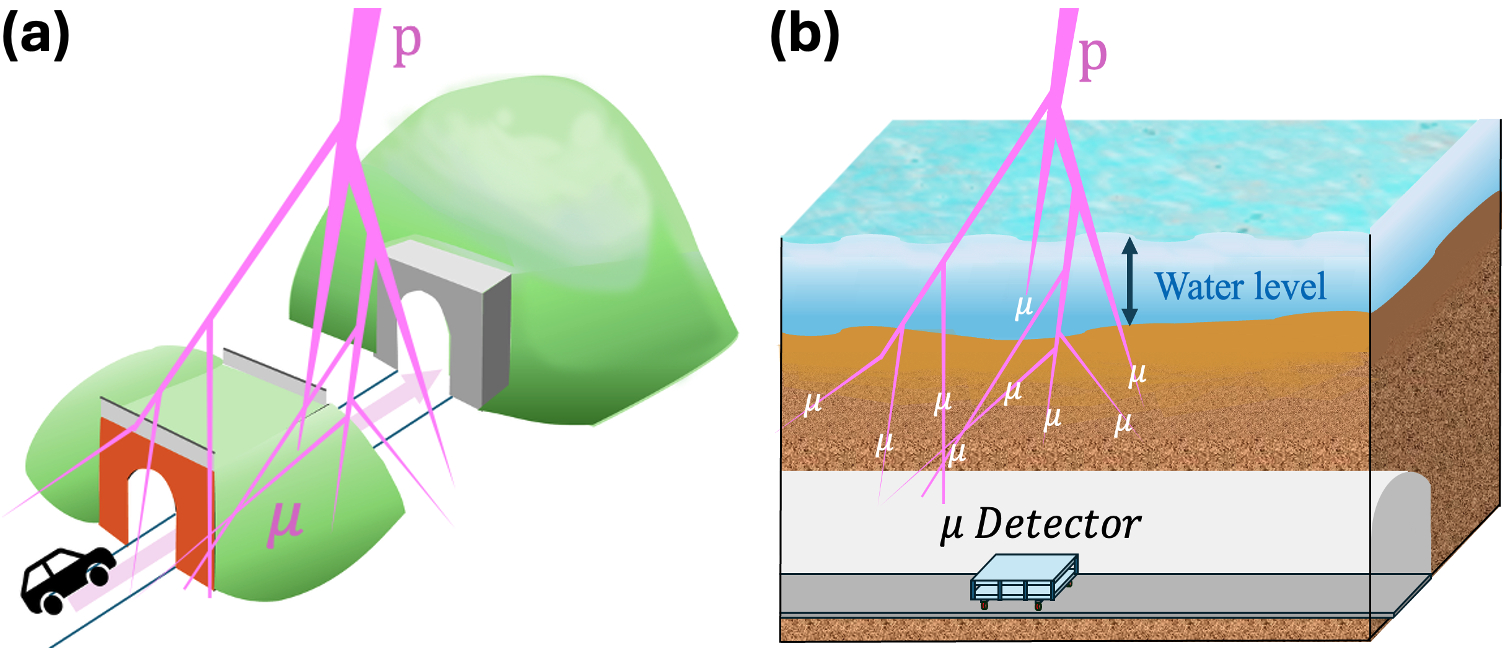}
  \end{center}
  \caption{
    Schematic diagrams of two approaches to dynamic muography.
    (a) Moving detector muography in the HAWL project.
    A vehicle mounted muon detector scans the overlying mountainous terrain
    while traveling through multiple tunnels at high speeds, and the muon rate data are transmitted in real time via the internet.
    (b) Time-varying-target muography in the HAWL upgrade project.
    The detector remains at a fixed location to measure changes in the muon rate induced by temporal variations
    in the seawater overburden above the undersea tunnel.
  }
  \label{hawl1}
\end{figure*}

\section{Introduction}
Cosmic-ray muons are produced by interactions of primary cosmic rays with atomic nuclei in the atmosphere~\cite{ParticleDataGroup:2024cfk}.
Due to their high penetrating power and unique energy-loss characteristics,  
variations in the transmitted muon rate can be used to investigate the internal density distributions of large structures and geological features~\cite{Bonechi2020,Muon_image}, such as cultural heritage structures and volcanoes~\cite{Procureur2023,Tioukov2019,Tanaka:2023ami}.
Muographic research has expanded to include a diverse range of targets, some of which change over time~\cite{Tanaka2014,Jourde2016,Tanaka2021,Khaw:2025tor}. 

Here, we present two approaches to dynamic muography, as shown in Fig.~\ref{hawl1}.
In the HAWL project, a portable detector was mounted on a vehicle and moved through tunnels beneath mountainous terrain~\cite{Seo:2024ijp}. 
This scan allowed a single detector to reconstruct the overburden distribution over a wide area.
In contrast, the HAWL upgrade (HAWLU) project aims to measure a target that changes over time using a detector fixed at a single location, and ocean tides provide an ideal natural setting for demonstrating this approach.
The west coast of Korea exhibits tidal level variations of approximately $\pm 4$~m~\cite{TideData2}, and the dominant semidiurnal period differs from the 24h cycle, helping to distinguish the tidal signal from diurnal environmental effects~\cite{Teague1998,Jung2013}.
The 6.9~km-long Boryeong Undersea Tunnel provides a stable underground environment by reducing the effects of local environmental variations~\cite{BoryeongTunnel2021,Korea-Tunnel}.
The HAWLU detector additionally performed an independent real-time moving scan to further validate its operational stability and measurement consistency.
Both HAWL and HAWLU use plastic scintillators~\cite{Eljen2023,Knoll2010}, with HAWLU incorporating a two plane configuration and direct SiPM coupling.

This work presents the detector development, event selection, and results of the two measurement approaches.
The tidal measurements are further supported by Fourier analysis, an evaluation of environmental effects, and simulations.

\section{Detector Development and Event Selection}
The HAWLU detector offers improvements over the original HAWL detector in detector configuration, SiPM coupling, and event reconstruction capability.
The HAWL detector used a single $600 \times 480 \times 30~\mathrm{mm}^3$ rectangular plastic scintillator plate
read out by 28 silicon photomultipliers (SiPMs) coupled to optical fibers~\cite{Kuraray2024,Hamamatsu2024MPPC}.
Eight SiPMs were arranged along each long side of the upper surface and six along each short side of the lower surface, with the optical fibers on the two surfaces arranged perpendicular to each other, enabling two dimensional hit-position reconstruction within the plate.
Each SiPM was mounted on a front-end electronics board,
which transmitted the signals to the data-acquisition (DAQ) system through a LAN cable.
Tyvek and aluminum foil were used as reflective and light-shielding layers. A stainless steel housing provided additional optical shielding and mechanical protection.

\begin{figure}[H]
  \centering
  \includegraphics[width=0.90\textwidth]{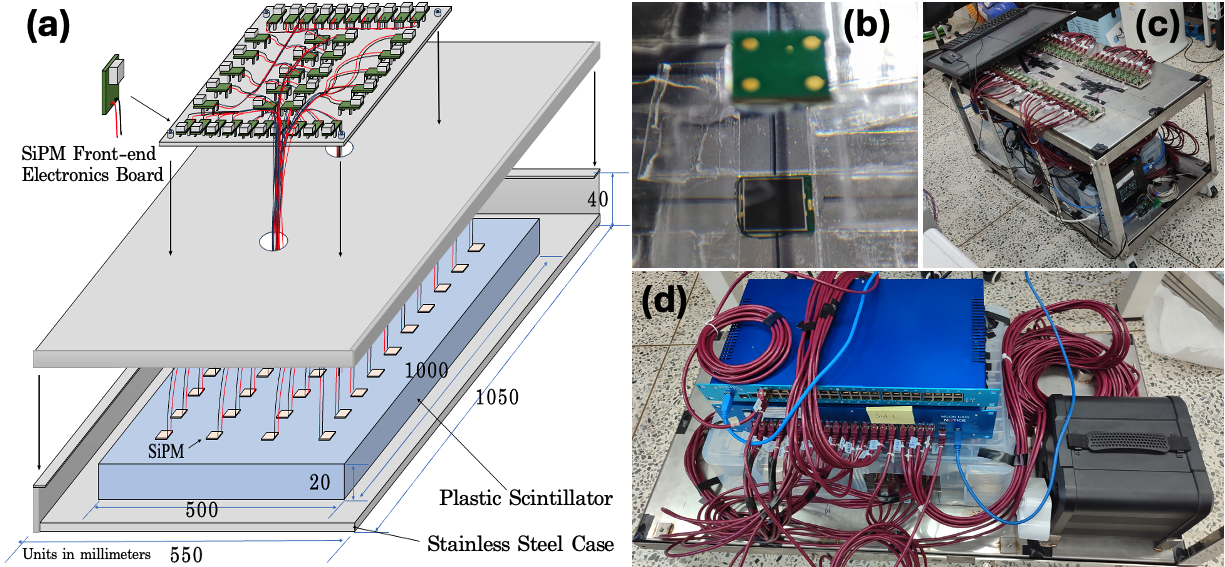}
  \caption{
    Design and assembly of the HAWLU detector.
    (a) Blueprint of a single detector panel.
    40 SiPMs are coupled to the plastic scintillator surface.
    The SiPMs are connected to the front-end electronics boards through thin signal wires.
    (b) Direct attachment of an SiPM to the plastic scintillator.
    An Enhanced Specular Reflector (ESR) film is attached to the lower surface of the scintillator, and an SiPM is bonded to the upper surface with optical cement.
    The reflected image of the SiPM indicates clean optical coupling.
    (c) Assembled detector.
    The detector is assembled as a single integrated unit and equipped with wheels for easy transport.
    The electronics required for data acquisition are located between the two detector panels.
    (d) Data acquisition electronics.
    The system includes two flash Analog-to-Digital converters, a trigger and clock board, a desktop computer, a battery, LAN cables, and a thermometer~\cite{Adhikari2018b}.
    The front-end electronics boards for the lower panel are positioned beneath the data-acquisition system.
  }
  \label{hawl2}
\end{figure}
The design and assembly of the HAWLU detector are shown in Fig.~\ref{hawl2}.
The HAWLU detector was constructed using two $1000 \times 500 \times 20~\mathrm{mm}^3$ plastic scintillator plates separated by 42 cm.
A total of 40 SiPMs are directly attached to the upper surface of each plate in a $10 \times 4$ grid, giving a total of 80 SiPMs.
The direct coupling of the SiPMs without optical fibers results in a more compact design and improved light collection. 
The SiPM arrangement was designed to provide spatial patterns in both light intensity and signal timing that enable effective reconstruction of the hit position in each detector plane.
The two plane configuration enables muon-trajectory determination from the reconstructed positions and improves muon–gamma discrimination through coincidence selection, thereby extending the measurable overburden range beyond that of HAWL.

\begin{figure*}[!htb]
  \begin{center}
      \includegraphics[width=0.99\textwidth]{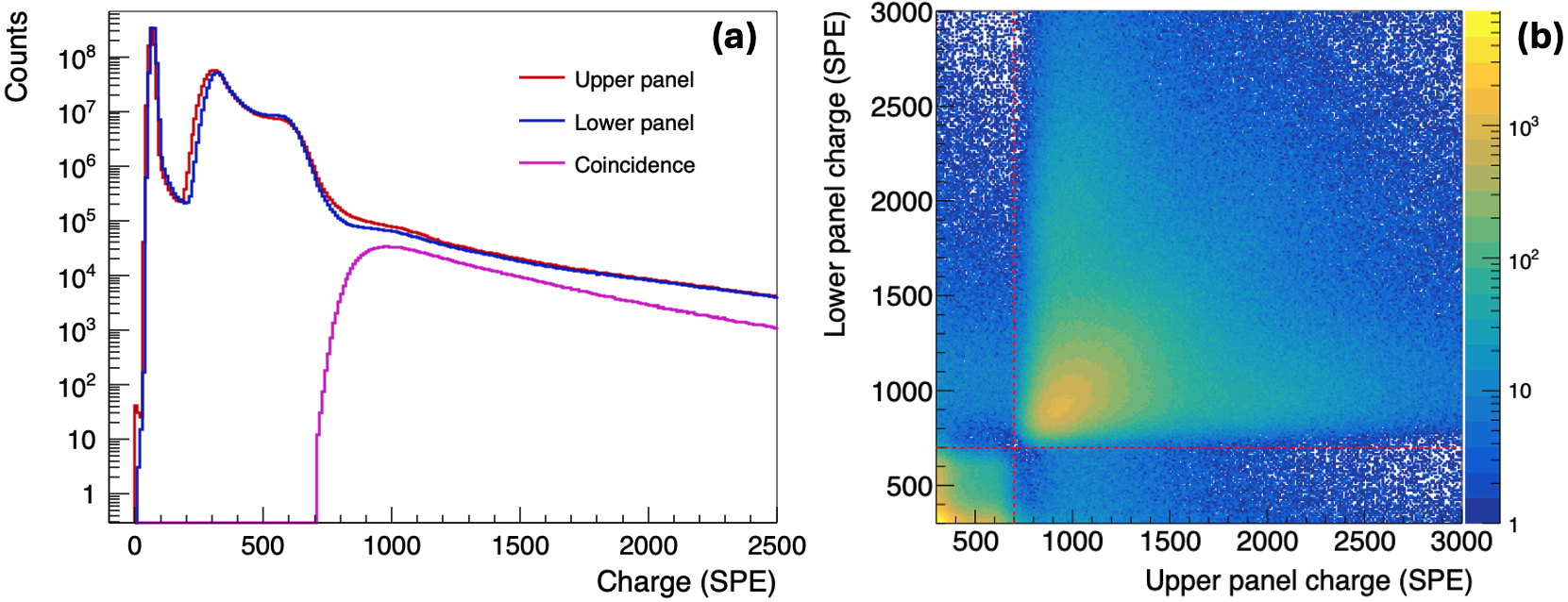}
  \end{center}
  \caption{   
    Charge spectra and coincidence event selection with the HAWLU detector.
    (a) Corrected charge spectra for the individual panels and coincidence-selected events.
    The red and blue curves show the total corrected charge measured in the upper and lower panels, respectively.
    The first structure at low charge reflects the threshold effect, whereas the subsequent spectral structures correspond to the Compton edges of gamma rays from $^{40}\mathrm{K}$ and $^{208}\mathrm{Tl}$ ~\cite{Be2004}.
    The magenta curve shows the mean of the total charges in the two panels, $(Q_{\mathrm{up}}+Q_{\mathrm{down}})/2$, for events satisfying the coincidence selection criteria.
    (b) Two dimensional distribution of the corrected charges measured in the upper and lower panels.
    The dashed lines indicate the 700 SPE thresholds applied to both panels, and events exceeding both thresholds were selected as muon candidates.
  }
  \label{hawl3}
\end{figure*}

An integrated charge signal exceeding 200~ADC counts within a 100~ns sliding window is defined as a hit, and the trigger condition requires hits in at least eight of the 80 channels within a 1000~ns window.
The charge measured by each SiPM was calibrated in units of single photoelectron (SPE) to account for differences in the channel responses.
Additionally, time dependent gain variations were corrected in each one hour interval by aligning the $^{208}\mathrm{Tl}$ Compton edge in the charge spectrum.
For each recorded event, the charge signals from the 40 SiPMs were summed separately for the upper and lower detector planes.
Muon candidates were selected by requiring coincident signals from the two planes with the total charge in each plane exceeding 700~SPE.
The two-plane event selection provided better separation between muon signals and the gamma-ray background than selection using a single plane (Fig.~\ref{hawl3}).

\section{Fast Moving Tunnel Wide Scan with HAWL}
\begin{figure*}[!htb]
  \begin{center}
      \includegraphics[width=0.99\textwidth]{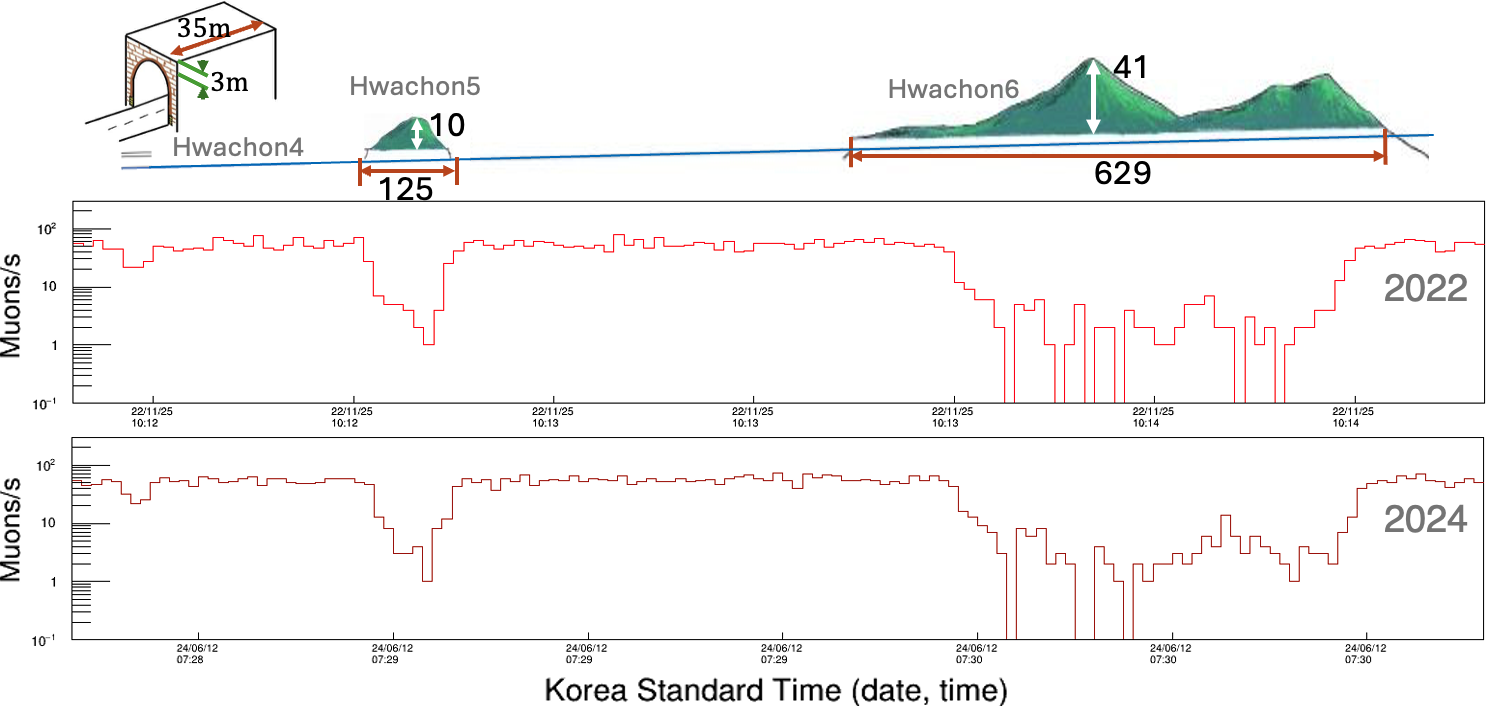}
  \end{center}
  \caption{
    Representative HAWL measurements.
    The upper schematic shows representative geometries of three tunnels along the Seoul–Yangyang Expressway.
    These examples differ in length and overburden thickness and include the animal overpass with the smallest overburden among all surveyed structures.
    The terrain profiles of Hwachon5 and Hwachon6 were derived from engineering drawings, whereas the Hwachon4 animal overpass is shown schematically.
    All dimensions in the upper schematic are given in meters.
    The middle and lower panels show the measured muon rates from Trip~1 (2022) and Trip~2 (2024), respectively.
    The horizontal axes indicate the measurement time in Korea Standard Time, and the vertical axes indicate the muon rate in counts/s.
  }
  \label{hawl4}
\end{figure*}

For the first time, the HAWL project successfully demonstrated portable muography over more than 100 km within a few hours.
During the measurement, the detector was placed in the trunk of a vehicle traveling along the Seoul–Yangyang Expressway at approximately 55~km/h.
Muon rate variations were recorded for all 47 tunnels.
The two measurements were conducted on November 25, 2022, and June 12, 2024, and the results from the two trips were consistent despite differences in the measurement conditions.
As shown in Fig.~\ref{hawl4}, HAWL was able to measure not only tunnel lengths and overburden thicknesses, but also detailed terrain features such as slopes, and even resolve an overburden thickness of only approximately 3 m above an animal overpass tunnel.

Tunnel lengths and overburden thicknesses were obtained by fitting the measured muon rate with an extended sigmoid function.
The sharp changes in the muon rate at the tunnel entrance and exit were useful for extracting tunnel parameters from the sigmoid fit.
The tunnel lengths were calculated using the fitted transit times and the vehicle speed, which was assumed to be the same for all tunnels.
The tunnel length reconstruction was evaluated using the fractional difference,
$\delta_L=(L_{\mathrm{measured}}-L_{\mathrm{true}})/L_{\mathrm{true}}\times100~\%$.
The tunnel lengths were reconstructed with an overall accuracy at the 6.0~\% level, as quantified by the fractional RMS difference.
The mean deviation of -1.7~\% indicates a small negative bias in the reconstructed tunnel lengths.
The detectable overburden range extended from 8 to 400 m.w.e., corresponding to approximately 3 to 150~m.
Gamma-ray contamination limited reliable measurements above 400 m.w.e.
The five tunnels near 10 m.w.e. deviated from the exponential attenuation model with a constant background because they corresponded to animal overpasses with geometries different from those of mountain tunnels.

\begin{figure*}[!htb]
  \begin{center}
      \includegraphics[width=0.8\textwidth]{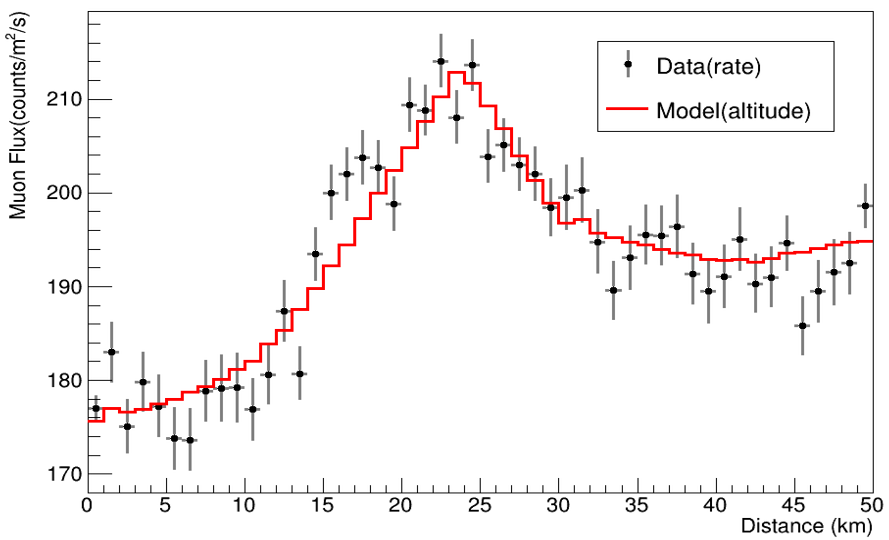}
  \end{center}
  \caption{
    Measured muon rate as a function of distance along the road.
    The HAWL detector traveled along the Guryongnyeong mountain road while measuring the open-sky muon flux and recording the vehicle speed and elevation.
    The measured muon flux followed the elevation dependence described by a model.
  }
  \label{hawl8}
\end{figure*}

This result was further supported by open-sky muon-flux measurements at Guryongnyeong Mountain, which is located between Seoul and Yangyang and has an elevation of $1{,}013~\mathrm{m}$.
The measured muon rate followed the overall elevation dependence, as shown in Fig.~\ref{hawl8}.
This trend validates the detector response.
A fit of the elevation model to the measured muon rates yielded $\chi^2/\mathrm{ndf} = 74.3/48$ with a p-value of 0.0088.

\section{Dynamic Tidal Modulation Measurement with HAWLU}
\begin{figure*}[!htb]
  \begin{center}
      \includegraphics[width=0.99\textwidth]{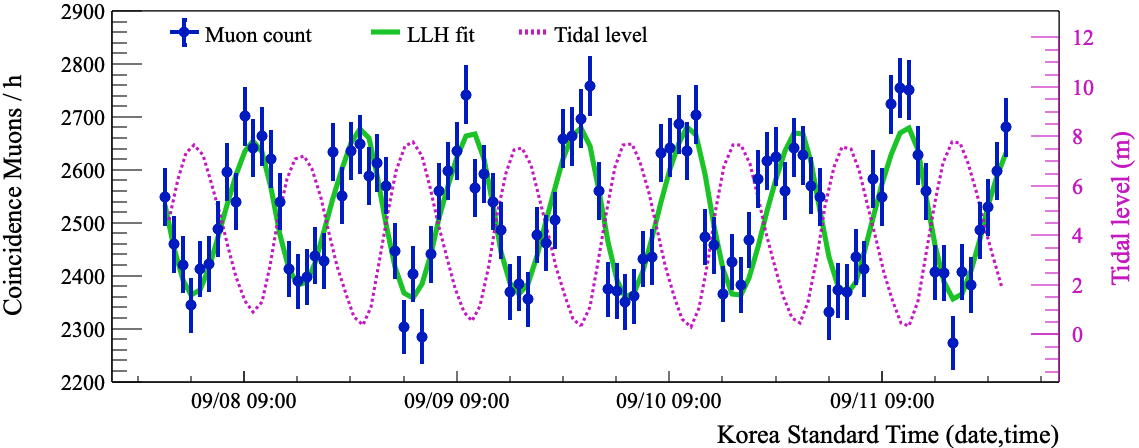}
  \end{center}
  \caption{
    Close-up of four days of HAWLU measurements around September 11, 2025, when the largest tidal range within this period occurred.
    The blue points indicate the measured muon counts and the pink dashed line shows the tidal level from the Boryeong tide gauge.
    The green solid line represents the best-fit model obtained using a Poisson likelihood fit to the full 31.6 day measurement period~\cite{BAKER1984437}.
  } 
  \label{hawl5}
\end{figure*}

Muons were measured for 31.6 days at a fixed location within the Boryeong Undersea Tunnel.
The detector operation site within the tunnel was selected to optimize sensitivity to tidal sea level variations, with a maximum seawater depth of 33.8 m and a minimum rock overburden thickness of 26.0 m~\cite{Korea-Tunnel2}.
The measurements were performed from August 11 to September 12, 2025, with a total data-taking time of 2,728,960 s, including a period when tidal ranges are among the largest of the year.
The measured muon rate decreased as the tidal level rose and increased as it fell, and the fitted model reproduced the observed periodic variations in the muon rate (Fig.~\ref{hawl5}).

The expected muon rate $\mu(t)$ at time $t$ was described by the following model
\begin{equation}
  \mu(t) = R_0 - k \Delta h(t - \Delta t),
    \label{eq:muon_tidal_model}
\end{equation}
where $\Delta h(t)$ denotes the deviation of the tidal level recorded at the Boryeong tide gauge from its mean over the measurement period.
$R_0$ is the mean hourly muon count, and the attenuation response coefficient $k$ corresponds to the decrease in the hourly muon count per meter of tidal-level rise.
$\Delta t$ represents the time shift of the muon rate variation relative to this station, which was the closest to the detector at a distance of 6.13~km.
The tidal level variations recorded at the Boryeong tide gauge were used as input to the model, and $R_0$, $k$, and $\Delta t$ were treated as free parameters in the fit.
The fit yielded $R_0 = 2514.64 \pm 1.82$~counts/h, $k = 43.21 \pm 0.95$~counts/h/m, and $\Delta t = -413.67 \pm 157.01$~s.
The muon rate response to a 1~m change in tidal level was expressed as a percentage of the mean muon rate and calculated as
\begin{equation}
  100\times\frac{k}{R_0} = 1.72~\%/m.
\end{equation}
A 1~m rise in tidal level corresponds to a 1.72~\% decrease in the muon rate relative to its mean value.
Despite this small muon rate response, the tidal modulation was clearly observed in the detector data.
The negative time shift suggests that tidal level variations above the detector occur earlier than those recorded at the Boryeong tide gauge.
The fit yielded $\chi^2/\mathrm{ndf}=768.1/755$ with a $p$-value of 0.36, and the RMS of the residuals was 50.5~counts/h.
These results show that the tidal-level model describes the observed muon rate variations well.
Assuming Poisson counting statistics, the expected statistical sensitivity per meter of tidal-level variation over a measurement time $T$ was estimated as
\begin{equation} 
    S(T) = \frac{k}{\sqrt{R_0}}{\sqrt{T}}
    = 0.86 \sqrt{\frac{T}{1~\mathrm{h}}}\,\sigma/\mathrm{m}.
\end{equation}
The statistical sensitivity increases with the square root of the measurement time.
Fitting the hourly muon rates over the full measurement period combines information from multiple time bins, allowing a more precise determination of the tidal response coefficient.

\begin{figure*}[!htb]
  \begin{center}
    \includegraphics[width=0.65\textwidth]{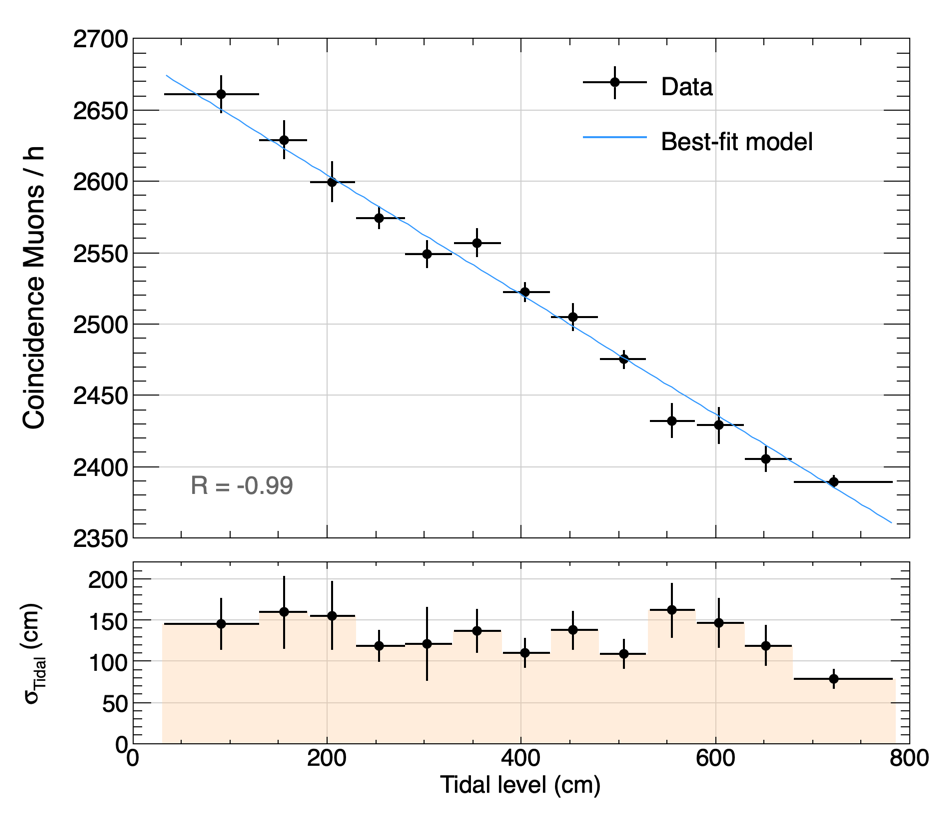}
  \end{center}
  \caption{
    Mean hourly coincidence muon counts in each tidal level interval and the uncertainty in the reconstructed tidal level.
    The black points show the mean hourly muon counts obtained from Gaussian fits to the count distributions in individual tidal level intervals.
    The horizontal error bars indicate the tidal level intervals, and the vertical error bars represent the uncertainties in the fitted means.
    The blue line shows a linear fit to the mean muon counts. 
    The uncertainty in the reconstructed hourly tidal level $\sigma_{\mathrm{tidal}}$ is shown in the lower panel and ranges from approximately 0.8 to 1.7~m across the measured tidal range.
  }
  \label{hawl6}
\end{figure*}

To evaluate whether the measurement retained meter-scale resolution across the tidal range, the data were divided into tidal-level intervals.
Within each narrow tidal-level interval, the variation in seawater overburden was small, allowing the expected hourly muon count to be treated as approximately constant.
Because the mean hourly muon count was sufficiently large at approximately 2515, the count distribution in each tidal-level interval was fitted with a Gaussian function.
The fitted mean represents the average hourly muon count in each tidal-level interval, and the fitted standard deviation quantifies the spread of the hourly counts.
As shown in Fig.~\ref{hawl6}, the fitted mean hourly muon count decreased with increasing tidal level, showing a strong anticorrelation with correlation coefficient $r=-0.99$.
A linear fit gave a slope of $-41.9 \pm 1.2$~counts/h/m, consistent in magnitude with $k$ obtained from the time-dependent fit.
The Gaussian width ($\sigma_\mu$) was converted into an uncertainty in the reconstructed water level using the fitted parameter $k$ according to
\begin{equation}
    \sigma_{\mathrm{tidal}} = \frac{\sigma_\mu}{|k|}.
\end{equation}

\section{Validation of the Tidal Measurement}
\begin{figure*}[!htb]
  \begin{center}
      \includegraphics[width=0.99\textwidth]{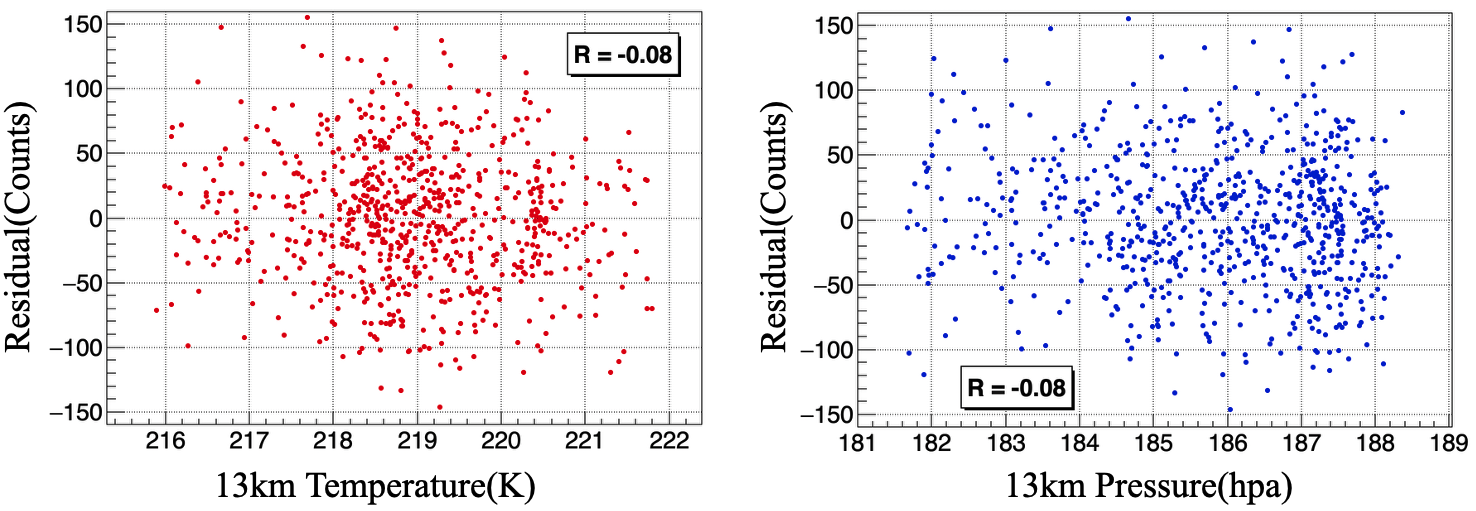}
  \end{center}
  \caption{
    Correlations between the fit residuals and atmospheric temperature(K) and pressure(hPa) at an altitude of 13~km~\cite{TP_Data,ERA5PressureLevels,Hersbach2020}.
    The residuals were obtained by subtracting the best-fit model from the measured muon counts in 1h bins.
  }
  \label{hawl7}
\end{figure*}

Muons are produced in the atmosphere, and atmospheric temperature and pressure can affect the measured muon count~\cite{COSINE-100:2020jml,Abdullrahman2025}.
Environmental effects were evaluated by examining correlations and adding an environmental response term to the fitting model.
The Pearson correlation coefficients between the residuals and atmospheric pressure or temperature were both $r=-0.08$, indicating negligible correlations with these environmental factors, as shown in Fig.~\ref{hawl7}.

With an additional environmental-response term included, the measured hourly muon counts were modeled as
\begin{equation}
\mu(t) = R_0 - k \Delta h(t - \Delta t) - \alpha_X X(t),
\label{eq:PT_term}
\end{equation}
where $X$ denotes either atmospheric temperature $T$ or atmospheric pressure $P$, and $\alpha_X$ indicates the corresponding response coefficient.
The fitted temperature and pressure response coefficients were consistent with zero, with respective 90~\% upper limits of 0.6~counts/h/$^\circ$C and 0.6~counts/h/hPa.
In contrast, the hourly muon counts showed a strong anticorrelation with the tidal level recorded at the tide gauge, with $r=-0.85$.

Furthermore, a Fast Fourier Transform (FFT)~\cite{d3ea2d52-5ab2-3128-8b80-efb85267295d,Harris1978} was applied as a frequency domain test to validate the observed tidal modulation.
After mean subtraction and application of a Hann window to the hourly data to reduce spectral leakage, the resulting spectrum showed peaks at $12.3 \pm 0.2$~hours and $16 \pm 8$~days, consistent with the 12.4~hours semidiurnal and 14.8~days spring-neap tidal cycles, respectively~\cite{Renault2022,KVALE20065}.
These results indicate that the observed muon modulation is dominated by seawater overburden variations rather than atmospheric temperature and pressure.

To examine the relative timing differences between the tide gauges and the detector site, the model was fitted separately using data from six nearby tide gauges~\cite{TideData}.
The fitted tidal time shifts had an average uncertainty of \(\pm 2.63\) min, with individual uncertainties ranging from 155 to 158 s across the six tide gauges.
The fitted time shift decreased with increasing tide-gauge latitude, yielding a correlation coefficient of \(r=-0.96\).
The small uncertainties in the fitted time shifts and their strong correlation with tide gauge latitude provide further support for the robustness of the main fit.

Moving spatial measurements were also performed to validate the overburden model and measurement stability.
The HAWLU detector performed ten moving scans through the Boryeong Undersea Tunnel on June 30, 2025.
An overburden model expressed in meters water equivalent (m.w.e.) was constructed from the tunnel engineering drawings by treating the seawater and seabed layers separately.
The model was used in a one dimensional simulation based on the Cosmic-Ray Shower Library (CRY), which generates secondary cosmic-ray particles, including muons~\cite{Hagmann:2007ziw,ART002186085}.
The simulation results were consistent with the combined muon data from the ten scans.

\section{Conclusion}
HAWL demonstrated real-time portable muography through wide area scans with a detector mounted in a moving vehicle.
The measured overburden ranged from 8 to 400~m.w.e., and the reconstructed tunnel lengths showed an accuracy of 6.0~\% with a bias of -1.7~\%.
Muon-flux measurements at elevations up to 1013 m along the Guryongnyeong mountain road also showed the altitude dependence, validating the reliability of HAWL.
HAWL was further developed into HAWLU, achieving a more compact configuration with position reconstruction capability. The detector consists of two plastic scintillator planes with direct SiPM coupling to increase the light yield.

HAWLU successfully resolved the time-dependent muon modulation caused by continuous changes in the seawater overburden, with the measured data well described by the tidal level model and the fitted response corresponding to a muon rate decrease of $43.2 \pm 1.0$~counts/h/m.
The fitted time shifts for the six tide gauges had a mean uncertainty of $\pm 2.63$~min and showed a strong negative correlation with tide gauge latitude ($r=-0.96$), further supporting the tidal level fit.
The FFT identified the semidiurnal and spring–neap tidal components, and residual-correlation analyses and likelihood fits including additional environmental response terms indicated negligible effects of atmospheric temperature and pressure.
Furthermore, the combined muon data from ten moving scans through the undersea tunnel were consistent with the simulation results.

The wide area spatial scanning demonstrated by HAWL and the precise time-resolved measurements achieved by HAWLU have potential applications in monitoring environmental changes associated with natural hazards such as tsunamis, typhoons, and sinkholes.

\putbib[uhawlJAIS]
\end{bibunit}

\section{Acknowledgments}
This work was supported by National Research Foundation of Korea~(NRF) grant funded
by the Korean government~(MSIT)(RS-2021-NR058750 and RS-2024-00438814).
\end{document}